\documentclass{aa}  

\usepackage{graphicx}
\usepackage{enumitem}
\usepackage{multirow}
\usepackage{booktabs}
\usepackage{array}
\usepackage{amsmath}
\usepackage{ragged2e}
\newcolumntype{L}[1]{>{\RaggedRight\arraybackslash}p{#1}}
\usepackage{txfonts}
\usepackage{xcolor}
\usepackage{comment}
\usepackage{subfig}
\usepackage[colorlinks, citecolor=blue, linkcolor=blue]{hyperref}
\usepackage[normalem]{ulem}

\providecommand{\gaia}{\textit{Gaia }}

\newcommand{\edit}[1]{\textbf{#1}}
\renewcommand{\edit}[1]{#1}

\begin{document} 

   \title{Characterisation of the Clouds' young stellar Bridge\\ using \gaia DR3} 
   %Some options: 
   %Identifying the young stellar Bridge between the Clouds in Gaia DR3
   %A young stellar Bridge sample in Gaia DR3/ The young stellar Bridge in Gaia DR3
   %Identifying a young Bridge of Gaia DR3 inter-Cloud stars
   %Characterising the young stellar Bridge (in Gaia DR3)?

   \subtitle{}

   \author{M. Sch\"olch \inst{1,2,3}$^,$\thanks{Email: marie.schoelch@fqa.ub.edu}
        \and Ó. Jiménez-Arranz \inst{4}
        \and M. Romero-Gómez\inst{1,2,3}
        \and X. Luri\inst{1,2,3}
        }

    \authorrunning{M. Sch\"olch et al.}

    \institute{Departament de Física Quàntica i Astrofísica (FQA), Universitat de Barcelona (UB), C Martí i Franquès, 1, 08028 Barcelona, Spain 
    \and
    Institut de Ciències del Cosmos (ICCUB), Universitat de Barcelona, Martí i Franquès 1, 08028 Barcelona, Spain 
    \and
    Institut d’Estudis Espacials de Catalunya (IEEC), C Gran Capità, 2-4, 08034 Barcelona, Spain
    \and
    Lund Observatory, Division of Astrophysics, Department of Physics, Lund University, Box 43, SE-22100, Lund, Sweden}

   \date{Received Month XX, 2026; accepted Month XX, 2026}

  \abstract
  {The interaction between the LMC and SMC (the Clouds) has resulted in prominent tidal features, including an extended bridge of gas and stars connecting the two galaxies. This Bridge has likely formed during the most recent interaction between the Clouds, about $150-250$ Myr ago. While some young stars observed in the Bridge have formed in-situ from the tidally stripped gas, stellar populations may also have been drawn out of the SMC during the tidal interaction.} 
  {We aim to identify a clean sample of likely Bridge stars in the region between the LMC and SMC using \gaia DR3 astrometric and photometric data combined with machine-learning techniques.}
  {We use the dimensionality-reduction algorithm UMAP to construct a training sample of young stars in the outskirts of the SMC and LMC. A neural network trained on this sample is then applied to \gaia sources in the inter-Cloud region to classify the stars and identify candidate Bridge members.}
  {We present and characterise a new sample of young candidate Bridge stars, selected from \gaia DR3. We investigate its spatial distribution, kinematic properties and colour-magnitude diagram and validate it using existing Bridge samples.}
  {The young stellar Bridge aligns well with HI gas, clusters, and cepheid samples, apart from a small offset near the LMC outer disc. We measure a Bridge length of $\sim 15$ kpc and the stars are travelling from the SMC to the LMC at a median tangential velocity of $\sim114\;\text{km s}^{-1}$. This implies a crossing time of $\sim 125$ Myr, which is within the timeframe of the last interaction of the Clouds and therefore supports tidal stripping as a possible formation scenario of the Bridge.} 

   \keywords{Galaxies: kinematics and dynamics - Magellanic Clouds - interactions - structure
               }

   \maketitle

%-------------------------------------------------------------------

\defcitealias{Jimenez-Arranz23a}{JA23}
\defcitealias{Jimenez-Arranz24a}{JA24}

\section{Introduction}
\label{sec:intro}
The LMC and SMC, collectively known as the Clouds or the MCs, are the closest interacting galaxies to the Milky Way (MW), and therefore well-suited for studying galaxy interactions at high spatial resolution. The galaxies' present-day morphology and kinematics are likely caused by their complex interaction history, as demonstrated in numerical simulations of the LMC-SMC-MW interaction \citep[e.g.][]{besla12,pardy18, Lucchini21, Jimenez-Arranz24a, Garver26}.
The LMC has a warped disc \citep[e.g.][]{vandermarel01, Olsen2002,Choi2018,Jimenez-Arranz25a, Oden25}, a tilted, off-centre, and possibly non-rotating bar \citep[e.g.][]{Zaritsky2004, Jimenez-Arranz24b, Rathore25a, Rathore25b, Jimenez-Arranz25b, Oden25}, and one single spiral arm \citep[e.g.][]{Ruiz-Lara20, Jimenez-Arranz25a}. The SMC is also in disequilibrium, with an irregular morphology, significant depth along the line of sight, and complex internal motions \citep[e.g.][]{stanimirovic04, subramanian-subramanian12, Zivick21, pardy18, Rathore25c}. All of these features are consistent with strong tidal interactions between the MCs and with the MW over the past few hundred Myr. % \citep[e.g.][]{pardy18, diaz-bekki12, Zivick21,  Rathore25b}.

Another result of the MCs' interaction is a bridge of gas and stars that lies between the two galaxies (hereafter simply `the Bridge', e.g. \citealp{misawa09, skowron14, carrera17, Zivick2019, GaiaLuri21}). The Bridge has been the target of many observational studies, to map both its gas and stellar components. However, due to its low stellar density compared to the high foreground contamination of MW field stars, the stellar population of the Bridge remains ill-defined.% \red{Rephrase? Old population contested, young population well established but no `complete' sample available (in my search at least..)?}  

The Bridge was first clearly identified in neutral hydrogen \citep{hindman63}. Early numerical simulations of the MCs were able to reproduce tidal features of the system, confirming that the presence of the gaseous Bridge is likely due to tidal stripping of gas from the SMC in a close encounter between the galaxies \citep{murai80, gardiner-noguchi96}. Then, optical surveys established that the Bridge also contains a prominent young stellar population. Concentrations of young main-sequence stars and cepheids prove that the Bridge was not formed entirely from tidal stripping, but also hosts in-situ star formation \citep[e.g.][]{harris07, Nidever11, Nidever13, Ficara26}. 

Using data from the Optical Gravitational Lensing Experiment (OGLE), \cite{skowron14} showed that young stellar populations form a continuous, albeit spatially clumpy connection between the two galaxies. They also reported the discovery of an overdensity of young stars roughly in the middle between the two Clouds, and called it the OGLE island. The distribution of the young stellar tracers is closely correlated with the highest HI column densities, which suggests that the Bridge's young stellar content likely formed in situ from tidally stripped gas. Searches for older and intermediate-age stellar populations continued in order to detect stars that could have been pulled from the SMC. As a part of the MAGellanic Inter-Cloud program (MAGIC), \cite{noel13} showed that a fraction of stars in the observed regions are intermediate age, and that there are even hints of an older population. \cite{bagheri13} also investigated 2MASS and WISE data and found a candidate population of older stars, which might have been tidally drawn out of the SMC. 
More recently, using \gaia DR1 data, \cite{Belokurov2017} reported the presence of an extended population of RR Lyrae stars tracing an old stellar bridge between the Clouds, suggesting that tidal interactions have also redistributed an ancient stellar component. This result was further supported by \cite{GaiaLuri21}, who used \gaia eDR3 astrometry to identify a diffuse population of old stars connecting the two galaxies. 

More recent studies such as the VIsible Soar photometry of star Clusters in tApii and Coxi HuguA (VISCACHA) survey analyse stellar clusters in the Bridge as a tracer of its formation history \citep{Oliveira23}. They found young, metal-rich clusters likely formed in-situ after the Bridge was formed, but also old clusters (500 Myr $-$ 6.8 Gyr) that may have been stripped from the SMC after formation. Using the STEP survey, \cite{Ficara26} recovered a star formation history of the Bridge that suggests primarily in-situ star formation from low-metallicity gas previously stripped from the SMC. They also confirm that the most recent interaction of the MCs was only a few tens of Myr before a peak in star formation they observe around 100 Myr ago.

Furthermore, the VISTA survey of the Magellanic Cloud system (VMC) in combination with \gaia DR2 has been used to trace the kinematics of the Bridge and confirms a flow of stars from the SMC to the LMC \citep{Schmidt20}.
Bridge candidates in this work were selected using colour–magnitude diagram criteria consistent with stellar populations of the Clouds and by applying parallax or distance constraints from \gaia-based StarHorse distances \citep{Queiroz18} to exclude nearby stars. The resulting sample allowed the authors to derive the median proper-motion field across the Bridge and to report coherent motion directed from the SMC toward the LMC. After the release of \gaia eDR3, \cite{GaiaLuri21} also showed that the stellar bridge between the Clouds can be clearly resolved in both density and proper motions.
These studies support a scenario in which the Bridge formed through tidal stripping during a recent interaction between the Clouds and demonstrates the capacity of \gaia astrometry to provide constraints on this encounter. Nevertheless, the selection of Bridge samples remains difficult, and studies based on \gaia DR3 are still limited.

% paragraph of aim and why it fills a gap?
As discussed in the previous paragraphs, the region between the LMC and the SMC has been studied extensively. Stellar ages and metallicities, as well as kinematic constraints, are central to modelling the Bridge’s formation by establishing a timeline of the interactions between the Clouds and by distinguishing between pure tidal stripping versus significant in-situ star formation scenarios. However, relatively few studies have taken full advantage of the \gaia coverage of the Clouds despite its unprecedented astrometric precision. This is largely due to the intrinsically low surface density of stars in the Bridge and the correspondingly low signal-to-noise of candidate members. In fact, a major challenge in identifying a clean sample of likely Bridge stars in the inter-Cloud region is the large number of foreground MW stars compared to the sparse population associated with the Bridge. In this paper, we address this difficulty by using machine-learning techniques to select a clean sample of likely Bridge stars from \gaia DR3 astrometric and photometric data.

This article is organised as follows: In Sect. \ref{sec:data}, we introduce the datasets used in this work. Sect. \ref{sec:method} describes the methods used to construct our Bridge sample, starting with an unsupervised dimensionality reduction followed by a neural network classifier. In Sect. \ref{sec:results}, we characterise the resulting Bridge sample, including its spatial distribution, proper motions, line-of-sight velocities, and colour-magnitude diagram (CMD). We compare the Bridge sample to previous studies and discuss the implications of our results in Sect. \ref{sec:discussion}. Finally, we summarise our work in Sect.\ref{sec:conclusions}.

%-------------------------------------------------------------------

\section{Data}
\label{sec:data}
The \gaia mission \citep{gaiadr2summary, gaiadr3summary} is designed to provide high-precision astrometry and photometry for sources across the sky, enabling detailed studies of stellar populations in the MW. Although it is optimised for the observation of MW stars, \gaia also includes extensive coverage of nearby galaxies, including the LMC and the SMC. In total, \gaia has observed on the order of 10–20 million stars associated with the MCs, yielding astrometry and photometry over their full spatial extent \citep{GaiaLuri21, Jimenez-Arranz23a, Jimenez-Arranz23b}. 

In this work, we use the complete LMC and SMC samples defined by \citet{Jimenez-Arranz23a} and \citet{Jimenez-Arranz23b}, respectively, which were constructed from \gaia astrometry and photometry. For these datasets, the authors improved the selection criteria of \cite{GaiaLuri21} by using a neural network classifier that identifies member stars of the Clouds while minimising foreground contamination. The complete samples include 12,108,920 %12,116,762 in paper
stars for the LMC and 2,172,427 stars for the SMC. Further details on the methodology and characterisation of the samples are available in the original works. 

Both galaxies' stellar contents are thereby well-characterised and help placing tidal features (such as the Bridge) in their dynamical context. Complementary to the LMC and SMC samples, in this work we aim to identify \gaia stars that belong to the Bridge. Therefore, we start by querying the \gaia DR3 catalogue in the region between the two galaxies. We replicate this `inter-Cloud' area defined in previous works \citep[e.g.][]{kerr54, harris07, bagheri13}, located between $70^{\circ}$ and $20^{\circ}$ in right ascension ($\alpha$), and between $-59^{\circ}$ and $-86^{\circ}$ in declination ($\delta$). After removing stars that lack colour information ($G_{\text{BP}}, G_{\text{RP}}$), we are left with 2,487,652 stars in our full \gaia inter-Cloud sample. %"Bridge-region sample" A density map of the region is shown in Fig. \ref{fig:umap2}).

The onset of the Bridge in the outskirts of the SMC is easily discernible in \gaia data, particularly in young stellar populations \citep[see for example Fig. A.7 in][]{GaiaLuri21}. %the young stellar population extending east of the main body of the SMC becomes very apparent. 
We therefore aim to define a dataset of young stars that are located at the edges of the Bridge, in the east wing of the SMC and in the western disc outskirts of the LMC. 
For \gaia eDR3, \cite{GaiaLuri21} presented samples of LMC and SMC stars, first making cuts in proper motion and then defining distinct areas in the CMD to identify stellar evolutionary phases. Among others, they identified two young main sequence populations of the MCs, 'young 1' with ages under 50 Myr, and `young 2' with ages between 50 and 400 Myr. We combine the CMD cuts of these two young stellar populations with the neural network classification of LMC and SMC stars \citep{Jimenez-Arranz23a, Jimenez-Arranz23b}, to yield a starting sample of likely Bridge stars.

In summary, we start by selecting LMC and SMC stars in the inter-Cloud region defined in Sect. \ref{sec:data} that are classified as `LMC' and `SMC' stars in the complete samples (with probability score $P > 0.01$) of \cite{Jimenez-Arranz23a} and \cite{Jimenez-Arranz23b}, respectively. Then we restrict this sample to the colour and magnitudes of the `young 1' and `young 2' populations defined by \cite{GaiaLuri21}. 
This initial sample of 655 stars serves as a training set to help us in the selection of the feature space to probe in the Uniform Manifold Approximation and Projection (UMAP, \citealp{umap}) distribution (see Sect \ref{subsec:umap}).

%-------------------------------------------------------------------
\section{Methods} %past tense!!
\label{sec:method}
In this section we describe how we constrained the full \gaia DR3 inter-Cloud sample from Sect. \ref{sec:data} to a clean sample of candidate Bridge stars. We started with UMAP, an unsupervised method to reduce the dimensionality of our dataset and isolate features of interest in the data (see Sect. \ref{subsec:umap}). We validated our feature selection with a clean sample before using it as an input for a neural network (NN) classifier in Sect. \ref{subsec:nn}. The supplemental NN classification ensures that our initial boundaries did not bias our selection.

\subsection{UMAP Bridge sample}
\label{subsec:umap}
In order to derive a sample of Bridge stars, we applied a dimensional reduction to our full inter-Cloud \gaia sample, described in Sect. \ref{sec:data}. We used UMAP to visualise our high-dimensional dataset in a low-dimensional projection. Since the map is based on continuous distributions of values, the data do not fall into distinct clusters. However, we can see underlying structures and similarities between data points in certain areas of the feature map, which helps us nonetheless to constrain regions of interest.

\begin{figure}
\includegraphics[width=0.95\columnwidth]{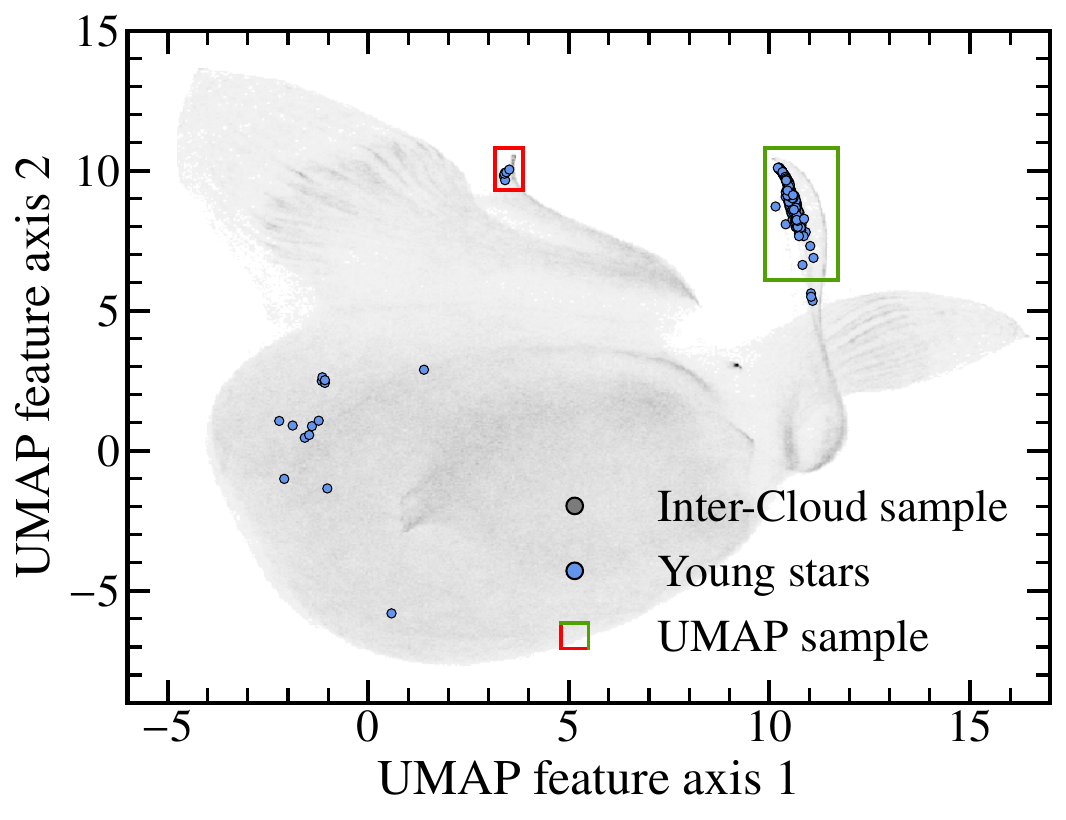}
\caption{Feature map from UMAP, showing the distribution of the full \gaia inter-Cloud sample, binned in grey. Overlaid is the sample of young LMC and SMC stars, plotted as blue circles. The red (green) frame highlights the region in the feature map that corresponds to the young stars of the LMC (SMC), which we select to constitute the UMAP Bridge sample (see Fig. \ref{fig:umap_pos}).}
\label{fig:umap1}
\end{figure}

The columns included in the UMAP feature reduction were the astrometric parameters position ($\alpha, \delta$), parallax ($\varpi$), proper motion ($\mu_{\alpha*}, \mu_{\delta}$), and the errors in parallax ($\sigma_\varpi$) and proper motion ($\sigma_{\mu_{\alpha*}}, \sigma_{\mu_{\delta}}$), as well as the photometric parameters colour ($G_{\text{BP}}, G_{\text{RP}}$) and apparent magnitude ($G$). We excluded the line-of-sight velocity ($V_{\text{los}}$) information due to its limited availability in \gaia data. \gaia DR3 includes line-of-sight velocities for $\sim 33$ million stars down to apparent magnitudes of $G=14$ mag \citep{katz23}, but only few of these stars are located in the inter-Cloud region. UMAP produces a feature map, presented in Fig. \ref{fig:umap1}. It shows the distribution of the full \gaia inter-Cloud sample \edit{(grey density map)} in a reduced feature space with two dimensions, here labelled UMAP feature axis 1 and UMAP feature axis 2. We emphasise that the axes of the UMAP embedding are arbitrary and do not carry intrinsic physical or interpretable meaning, as the method preserves relative neighbourhood structure rather than projecting data onto meaningful coordinate directions. 

To select the region in the feature map that represents the young Bridge population, we embedded our labelled young LMC and SMC samples from Sect. \ref{sec:data} on the same feature map. We can see an overdensity of this sample in the top right region of the map with coordinates (12.0, 9.0), highlighted by the green frame.  We note a second clustering of stars at the top of the feature map, at (3.0, 10.0), marked by a smaller red rectangle. Plotting the spatial distribution of the stellar populations in these two regions (see Fig. \ref{fig:umap_pos}) revealed that they correspond to young SMC stars in the onset of the Bridge (green frame), and young LMC stars in the LMC outskirts (red frame). %(Add Bica cluster validation and Be stars?)
We further inspected the parameters of stars located in other areas of the feature space, particularly the outliers of the young star sample, scattered around (-1,0). These \edit{stars' on-sky locations are} in the LMC and SMC haloes \edit{and they} exhibit large scatter in their proper motions. \edit{They also have \gaia colours $G_{\text{BP}} - G_{\text{RP}} \sim 0.5-1.5$ for a range of different apparent magnitudes, which results in a vertical distribution in the CMD that is} consistent with MW stars \edit{\citep[see for example Fig. 6 in][]{Jimenez-Arranz23a}}. We thereby concluded that these stars are MW contamination, and rejected these regions in the feature map. 
In order to prioritise purity in our training sample, we only selected \gaia stars in the feature map that overlap with the LMC and SMC young stellar samples, \edit{shown in the red and green rectangles}. This resulting `clean' Bridge sample from UMAP contains 18,788 stars.

\subsection{Neural network classifier}
\label{subsec:nn}
The labelled Bridge sample from the UMAP dimensional reduction described in Sect. \ref{subsec:umap} is by definition conservative and limited by inconclusive boundaries drawn on the feature map. To search for a more complete sample of Bridge stars, we therefore used a supervised neural network classifier on the \gaia DR3 data, with the conservative UMAP Bridge sample as a training set. The Bridge is very sparsely populated in comparison to the much larger population of field stars. In order to train the classifier to differentiate between the two populations, we supplemented the UMAP Bridge sample with MW field stars. We added 187,880 random stars from \gaia data in the inter-Cloud region, excluding specifically the stars that were in the UMAP selection as the Bridge. This gives us a 1:10 ratio of Bridge stars to MW stars.

The neural network is a fully connected model built for binary classification. It has two hidden layers with 64 and 32 neurons, respectively, both using the Rectified Linear Unit (ReLU) activation function. The output layer has a single neuron with a sigmoid activation and returns a continuous distribution of scores $S$ between 0 and 1 to predict whether a star is part of the Bridge ($S \sim 1$) or not ($S \sim 0$).
The model was trained using the Adaptive Moment Estimation (Adam) optimiser \citep{adamoptimisation} and binary cross-entropy loss \citep{crossentropylossfunctions}.
We divided our data into 80\% training and 20\% validation sets and trained over 30 epochs with a batch size of 32. 

\begin{figure}
\includegraphics[width=0.95\columnwidth]{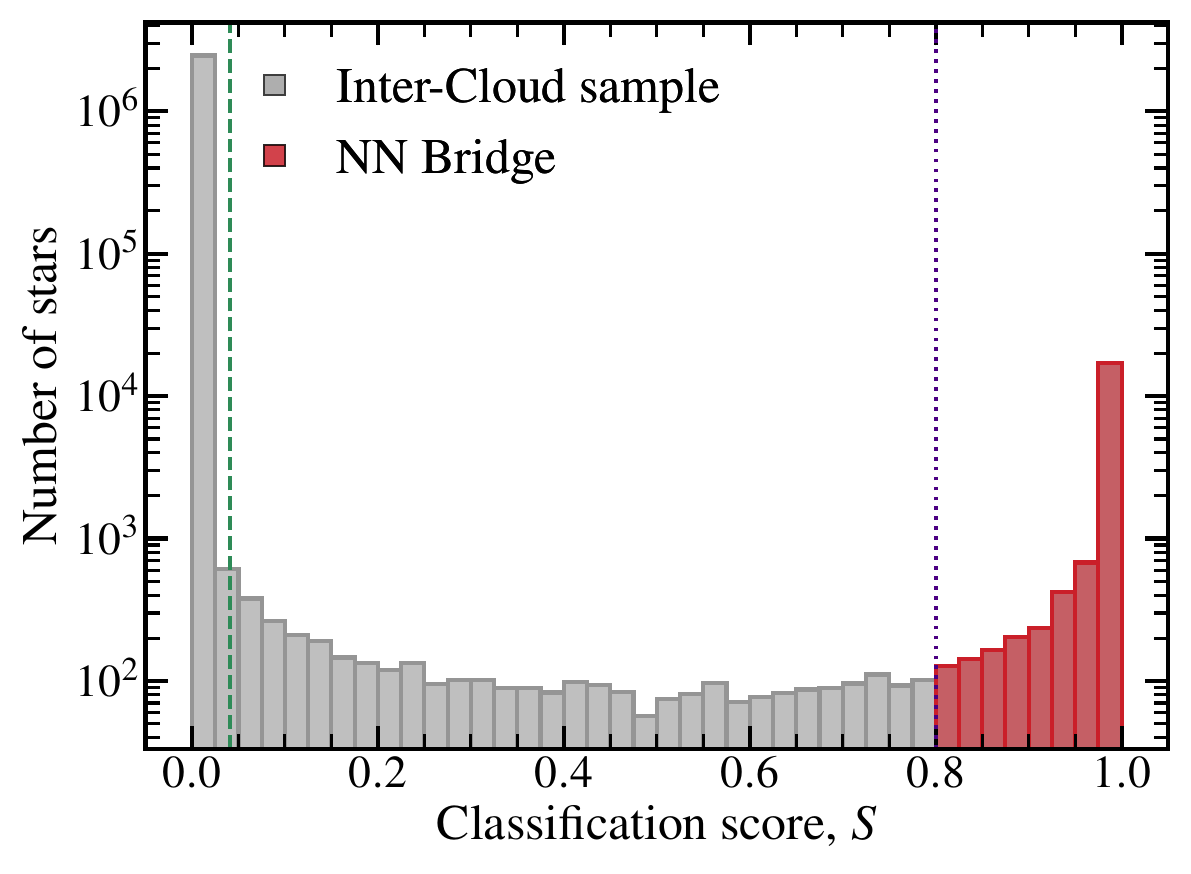}
\caption{The distribution of classification scores of the full \gaia DR3 sample in the inter-Cloud region (grey bins), and the NN sample (red bins). Note that the number of stars is shown on a log scale. The threshold of Youden’s J statistic $S_\text{cut} = 0.041$ is displayed in green (dashed line), and the chosen threshold of the final sample, $S_\text{cut}=0.8$, is shown in purple (dotted line).}
\label{fig:score}
\end{figure}

After training is complete, we have a continuous classifier. Since the aim is to have a binary classification of the \gaia DR3 stars, we need to determine the optimal score threshold between the two target classes. The ROC curve shows the balance between completeness and accuracy in the classifier's predictions (see top panel of Fig.~\ref{fig:roc_pr}). Since we have an imbalanced sample with a 1:10 ratio of Bridge to MW stars, we also show the precision-recall curve (see bottom panel of Fig.~\ref{fig:roc_pr}). We computed Youden’s J statistic, which marks the precise threshold that maximises the distance between the ROC curve and the `chance line' of a random classifier, a score of $S_\text{cut} = 0.041$. This score is a recommendation based only on the training sample, to achieve a trade-off between completeness and purity. However, in Fig. \ref{fig:score} we show that the full distribution of scores $S$ for our sample is strongly bimodal, and the large peaks of stars with classification scores of $S \sim 0$ and $S \sim 1$ dominate the low number counts in-between. As a result, we can choose various scores as thresholds for our sample while its size and properties remain consistent. In order to compromise between high purity and retaining a large number of stars, we chose a threshold of $S_\text{cut}=0.8$ for our final sample presented in this paper. Comparing the line-of-sight velocity distributions of the $S_\text{cut}=0.8$ sample in Sect. \ref{subsec:rv} with the $S_\text{cut}=0.041$ sample in Fig. \ref{fig:rv041} confirms that raising the threshold to $S_\text{cut}=0.8$ removes some contaminating MW sources from the sample.

To gain a better understanding of the machine learning model, we used a software called SHapley Additive exPlanations \citep[SHAP,][]{SHAP}. The SHAP evaluation plot in Fig. \ref{fig:shap} gives us more information about the NN classification. It sorts the input columns so that the most influential feature of the classification, in this case the \gaia colour $G_\text{BP}-G_\text{RP}$, is at the top. Every point in the plot represents a star, coloured by its feature value. For example, high declinations are coloured pink, and low declinations are coloured blue. 
The location of each star to the right or the left of the $x=0$ line represents whether the predicted class of the star is part of the Bridge (right side), or part of the MW background (left side). The neural network strongly reflects the input training sample described in Sect. \ref{subsec:umap} (for further discussion of its limitations, see Sect. \ref{subsec:lims}).

\subsection{Positional cut}
Applying the NN to the inter-Cloud \gaia data yielded a clean sample of 19,010 stars (with a score of $S>0.8$), hereafter called the NN Bridge sample. After inspecting the distribution of NN Bridge stars, we noted an overdensity located in the outskirts of the LMC. We therefore introduced a positional cut on the sample to exclude the stars belonging to a young LMC disc population (see Fig. \ref{fig:poscut}). This reduced the number of stars in the NN Bridge sample from 19,010 to 12,864.

%-------------------------------------------------------------------
\section{Characterisation of the NN Bridge sample}
\label{sec:results}
In this section, we characterise the NN Bridge sample of 12,864 stars derived in Sect.~\ref{sec:method}. We evaluate its position, proper motions, line-of-sight-velocities, and CMD. Lastly, we also discuss limitations to our Bridge sample. 

\subsection{Position}
\label{subsec:pos}
\begin{figure}
\includegraphics[width=0.95\columnwidth]{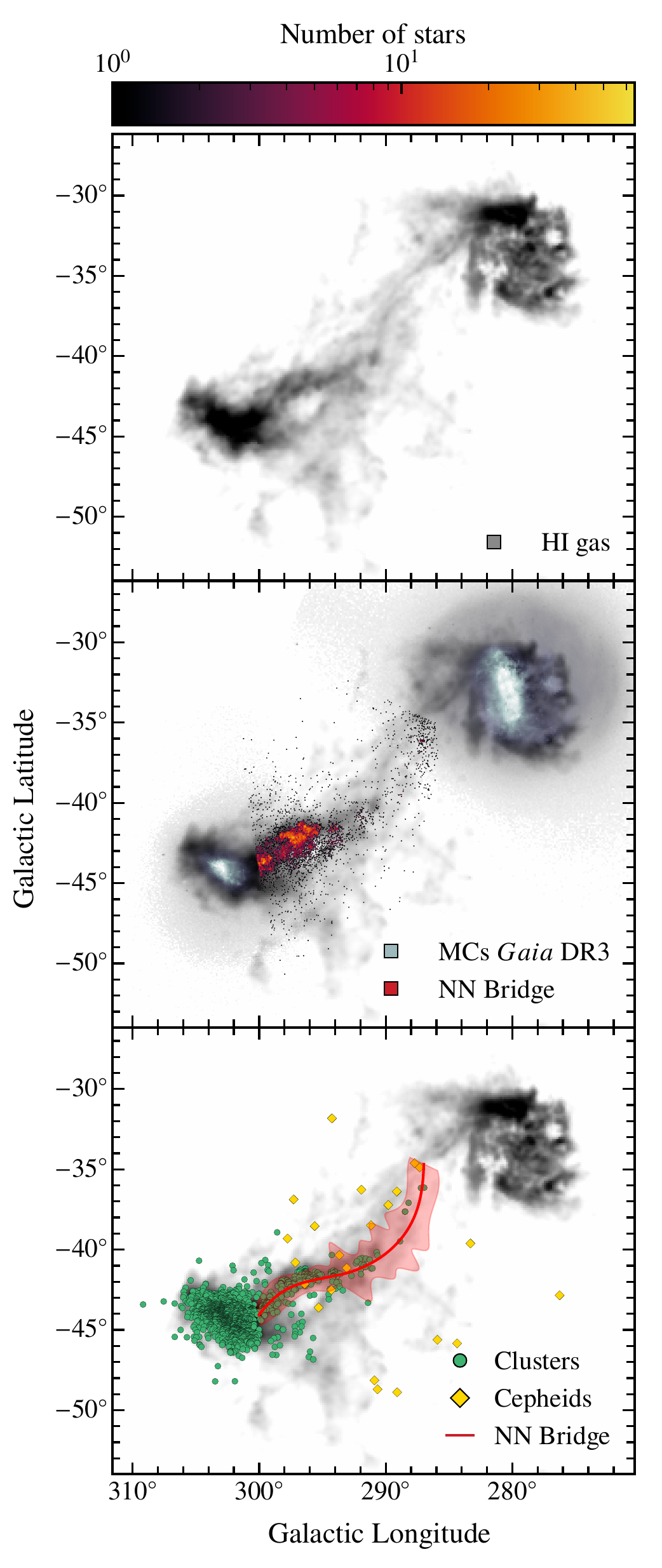}
\caption{Top panel: Map of intermediate- and high-velocity neutral hydrogen from the Galactic All-Sky Survey (GASS,  \citealp{GASSI, GASSIII}). The map also constitutes the background of the middle and bottom panels. 
Middle panel: Density plot of the NN Bridge sample, where bins with two stars or more are shown in shades of red and individual stars as a scatter plot in black. Density maps of the LMC and SMC complete samples from \cite{Jimenez-Arranz23a, Jimenez-Arranz23b}, respectively. Brighter colours correspond to higher density regions. % on top of the HI distribution. 
Bottom panel: Ridge line tracing the maximum density of the NN Bridge sample (red line) and its $\pm 1 \sigma$ dispersion (shaded red region). Scatter plots of the 2741 clusters in the SMC and the Bridge from the \cite{Bica20} catalogue (green points), and the 23 cepheids from \cite{jacyszyn-dobrzeniecka16} (yellow squares).} % All are shown on top of the map of neutral hydrogen from GASS.}
\label{fig:pos}
\end{figure}

In Fig. \ref{fig:pos} we present a 2D density map of the NN Bridge sample (middle panel) and a spline fit of the NN Bridge sample (bottom panel) in the galactic reference frame. We also show the SMC (middle panel, left) and the LMC (middle panel, right) density maps from \cite{Jimenez-Arranz23b, Jimenez-Arranz23a} as references to the Bridge's location. We compare the NN Bridge sample to a neutral hydrogen map from the Galactic All-Sky Survey \citep[GASS,][]{GASSI, GASSIII}, shown in the top panel and the backgrounds of the middle and bottom panels of Fig. \ref{fig:pos}. For the gas density maps, we present only intermediate- and high-velocity gas, since it corresponds to the Clouds.

The NN Bridge sample shows a clear trail of stars from the eastern SMC extending towards the LMC, with its highest density close to the SMC. Around the midpoint between the two galaxies, the Bridge stars become very sparse, but a continuation of the trail can still be seen even in the outskirts of the LMC disc. To better define the trail of young stars between the MCs, we fitted a smooth spline to the maximum density of the NN Bridge sample. In the bottom panel of Fig. \ref{fig:pos}, we show the ridge line tracing the highest density in red. The shaded red regions around it indicate the $\pm 1 \sigma$ dispersion of the sample, computed in small bins perpendicular to the spline. Using this dispersion, we determined the median width of the Bridge, $\sim 1 \pm 0.5 ^\circ$. %$\sim 1.01 \pm 0.41 ^\circ$. 
To obtain the length of the Bridge, we computed the arc length of the spline, $ \sim 15 ^\circ$, corresponding to a path length of $\sim15$ kpc at a distance estimated between 50 kpc and 62 kpc.

Close to the SMC, the position of the Bridge is very well aligned with the HI map. Approaching the LMC, the Bridge stellar sample no longer overlaps perfectly with the neutral hydrogen map. The misalignment between the stellar and gaseous Bridge is very apparent in the bottom panel, where the Bridge spline is offset by of $\sim 1^\circ$ from the neutral hydrogen. 

\subsection{Proper motions}
\label{subsec:pm}

\begin{figure}
\includegraphics[width=0.95\columnwidth]{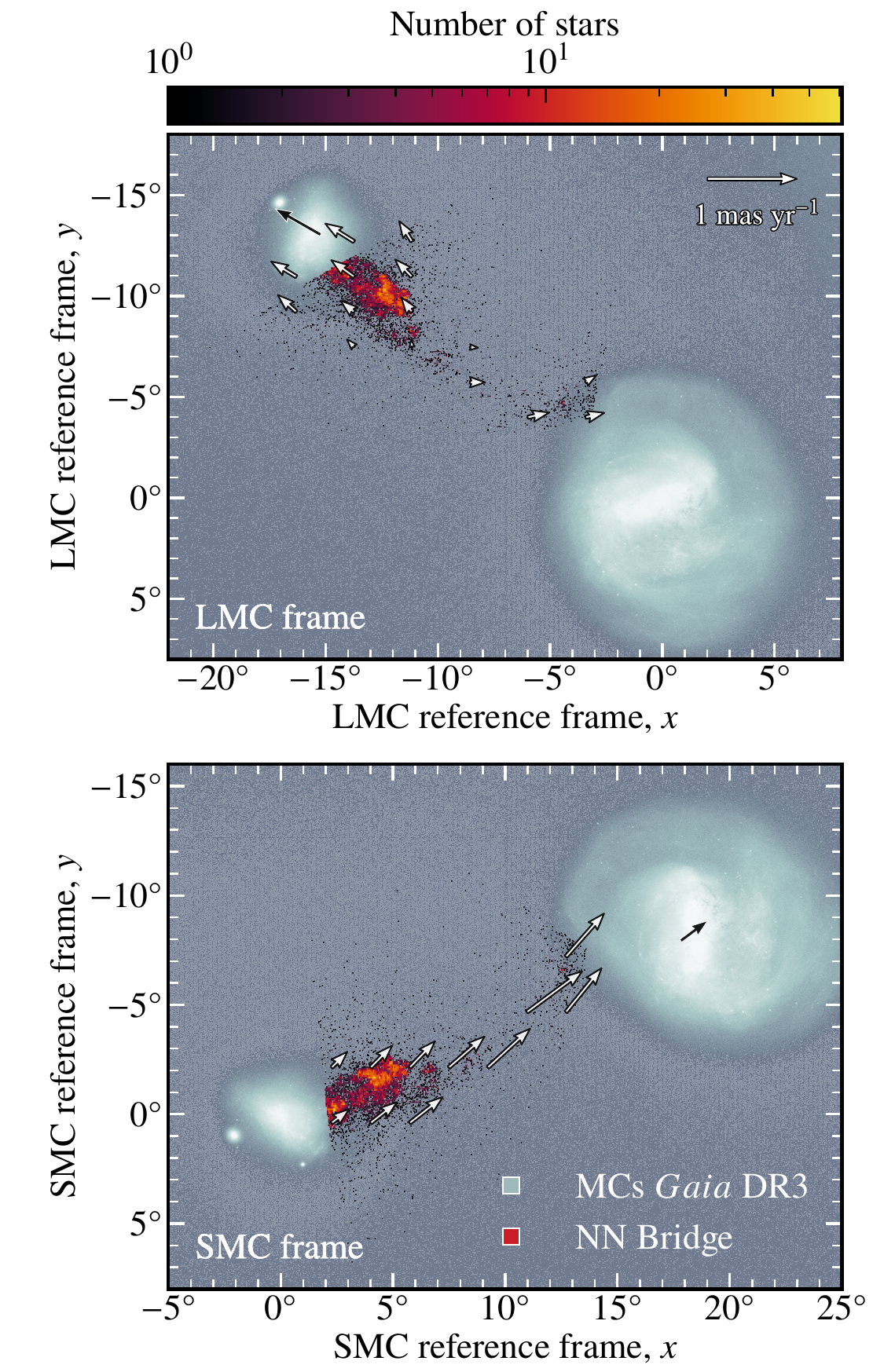}
\caption{Density maps of the NN Bridge sample, plotted on two-dimensional histograms of the \gaia DR3 sample in the region. The top (bottom) plot is in the LMC (SMC) orthographic reference frame, and shows the NN Bridge sample's proper motions with respect to the LMC (SMC) systemic motion with white arrows, while the motion of the SMC (LMC) is indicated with the black arrow.}% The bottom plot shows the NN Bridge sample and its proper motions with respect to the SMC centre position and systemic motion.}
\label{fig:pm}
\end{figure}

To visualise the proper motions of the Bridge, we compute the velocities with respect to the LMC and the SMC. For this purpose, we define orthographic projections of the heliocentric coordinates provided by \gaia DR3, as presented in equations (1)-(3) of \cite{GaiaLuri21}. The orthographic reference frames are created with respect to chosen centre coordinates. In this work, we create two reference frames, one centred on the LMC centre coordinates $(\alpha, \delta) = (81.28^\circ, -69.78^\circ)$ and the other on the SMC centre coordinates $(\alpha, \delta) = (12.80^\circ, -73.15^\circ)$. Then we compute the velocities in the respective reference frames and subtract the galaxies' systemic motions $(\mu_{\alpha*}, \mu_{\delta})= (1.871 , 0.391)\;\text{mas yr}^{-1}$ for the LMC and $(\mu_{\alpha*}, \mu_{\delta})= (1.858 , 0.385)\;\text{mas yr}^{-1}$ for the SMC. In Fig. \ref{fig:pm} we show the NN Bridge sample positions and velocities in the LMC reference frame (top panel), and the SMC reference frame (bottom panel). 

In the LMC reference frame (top panel), we note the clear trail of stars moving towards the LMC from the midpoint of the Bridge. On the other hand, stars near the SMC exhibit motion away from the LMC, with increasing magnitude closer to the main body of the SMC. This is because the SMC's systemic motion is slower than that of the LMC, so in the LMC reference frame, the SMC is moving away from the LMC, as indicated by the black arrow. As a result, the proper motions of the Bridge stars near the SMC centre are actually dominated by the bulk motion of the SMC and not by the gravitational pull of the LMC.

In the SMC reference frame (bottom panel of Fig. \ref{fig:pm}), we can clearly see the systemic motion of the LMC moving away from the SMC (black arrow). We also observe the general motion of all NN Bridge stars towards the LMC. 
We compute a median tangential velocity of $\sim114\;\text{km s}^{-1}$ for stars moving from the SMC toward the LMC.
%median tangential velocity: 114.27214677093642

\subsection{Line-of-sight velocities}
\label{subsec:rv}

\begin{figure}
\includegraphics[width=0.95\columnwidth]{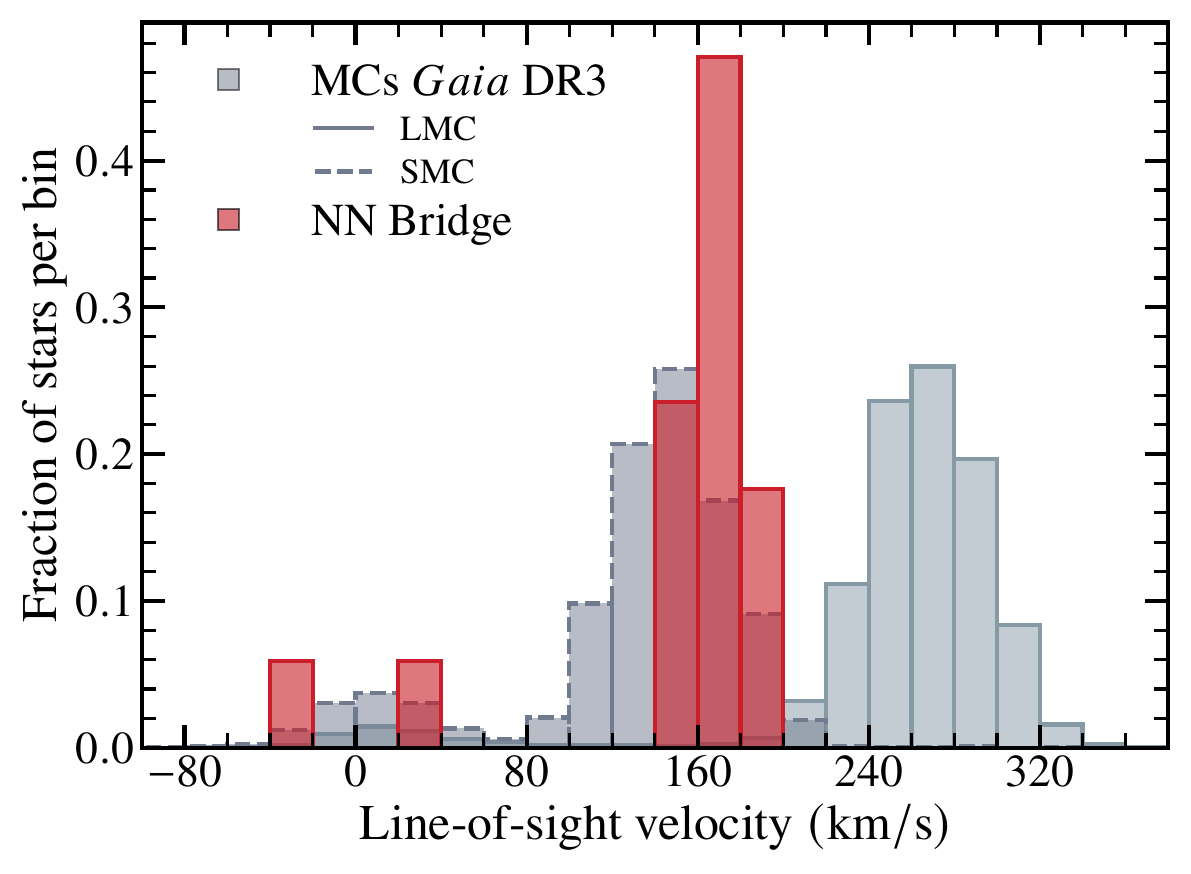}
\caption{Distributions of available line-of-sight velocities from \gaia DR3, for the LMC (grey bins, solid lines), SMC (grey bins, dashed lines), and NN Bridge (red bins) samples. The number of stars is normalised, so the vertical axis represents the fraction of stars per bin.}
\label{fig:rv}
\end{figure}

Since line-of-sight velocities were excluded from the classifier, they provide an independent validation of the purity of the NN Bridge sample. However, only a very limited sample of the NN Bridge stars has line-of-sight velocities available in the \gaia data. These 17 stars are also the brightest of the sample with a maximum of 14.48 in \gaia $G$ apparent magnitude. \edit{Despite the small number of stars with this velocity information available, i}n Fig. \ref{fig:rv} we show the distribution of line-of-sight velocities for the LMC and SMC complete samples (grey bins, solid and dashed lines, respectively) compared to the NN Bridge sample (red bins). The number of stars in each sample is normalised to represent the fraction of stars in each line-of-sight velocity bin. The line-of-sight velocities of the Bridge sample lie mainly between $140 - 200 \;\text{km s}^{-1}$, between the MC's individual distributions. They are generally consistent with the SMC velocity range which is a bit broader ($\sim 80 - 220 \; \text{km s}^{-1}$), but skew to larger velocities towards the LMC distribution ($\sim 180 - 340 \; \text{km s}^{-1}$). In all samples, a subset of sources shows velocities below $80 \; \text{km s}^{-1}$, likely reflecting contamination from MW foreground stars. We do not remove these likely MW stars from the NN Bridge sample since the number of stars that have line-of-sight velocities available is already very limited. Nevertheless, our threshold of $S_\text{cut}=0.8$ in the NN Bridge sample clearly already removed some of the MW contamination. This is evident by comparison with the line-of-sight velocity plot of the $S=0.041$ sample in Fig. \ref{fig:rv041}, which was the Youden's J statistic threshold to balance purity and completeness.

\subsection{Colour-magnitude diagram}
\label{subsec:cmd}

\begin{figure}
\includegraphics[width=0.95\columnwidth]{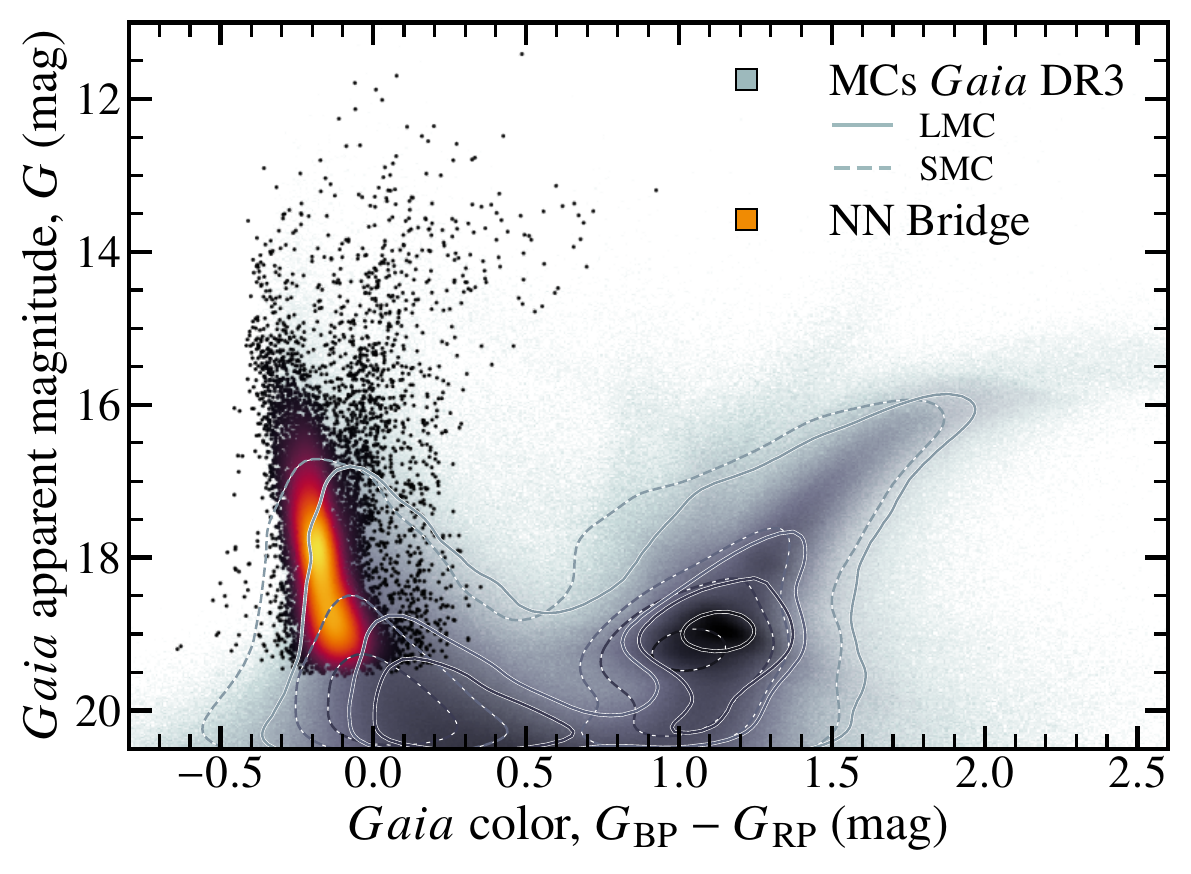}
\caption{CMD of the NN Bridge sample (black scatter with brighter colours for higher density), overplotted on the CMDs of the LMC and SMC complete samples  (grey density maps) from \cite{Jimenez-Arranz23a, Jimenez-Arranz23b}, respectively. To highlight the features of the MC's CMDs, we also show their contours in solid lines (LMC) and dashed lines (SMC).}
\label{fig:cmd}
\end{figure}

In Fig. \ref{fig:cmd}, we investigate the photometric characteristics of the NN Bridge sample by comparing its CMD to those of the MCs. We plot the \gaia apparent $G$ magnitudes as a function of the \gaia colours, $G_{\text{BP}} - G_{\text{RP}}$. The density map of the LMC and SMC CMDs are shown with added contours in solid and dashed lines, respectively. This way we can highlight features in the CMDs of the MCs \edit{which correspond to different evolutionary phases defined in the polygon regions of \cite{GaiaLuri21}. We note for example} the red clump, around $G_{\text{BP}} - G_{\text{RP}} \sim 1.2$ and $G \sim 19$, the red giant branch extending up from the red clump, and young populations on the blue end of the diagram. 

The colour-magnitude distribution of the NN Bridge stars aligns very closely with that of the SMC. The majority of the stars in the sample are clearly part of the young population, with an overdensity around $G_{\text{BP}} - G_{\text{RP}} \sim -0.1$ and $G \sim 18-19$. This is a direct result of our training sample, which is based on the young stellar populations defined in \cite{GaiaLuri21}. We also see this reflected in the SHAP plot (Fig.\ref{fig:shap}), which presents the colour as the predominant feature in the classification.

\subsection{Limitations}
\label{subsec:lims}
By the definition of our training sample, we have limited our study to finding only young stars in the Bridge. Since older and redder stars show a significant overlap in the CMD with the much more numerous MW foreground stars, we would need a very reliable training sample to include this distinction in the NN. We leave the search for older stellar populations in the Bridge for future work. %with a reliable training sample of old Bridge stars.

\section{Tidal stripping or in-situ formation}
%\section{Discussion}
\label{sec:discussion}

%\subsection{Tidal stripping or in-situ star formation?}
The SMC's tidal disruption has caused a trail of gas and stars originating from its eastern wing. In Fig.~\ref{fig:pos} we note the alignment between the ridge line of the NN Bridge sample (solid red line) and the neutral HI gas stripped from the SMC (grey density map). However, near the LMC, the spline does not align completely with the HI gas. This can likely be explained by the different mechanisms of tidal stripping for gas and stars. While gas stripped from the SMC may have been displaced by ram pressure and other interactions with MW and LMC halo gas, the stars' paths are largely influenced by the gravitational potential of the LMC.

%Reiterate how NN Bridge aligns with other bridge samples.. OGLE? Ficara?
In the bottom panel of Fig. \ref{fig:pos}, we compare the spatial distribution of the clusters and associations in the SMC and the Bridge published in the \cite{Bica20} catalogue (green scatter points) with the ridge line of the NN Bridge sample (red solid line). The NN Bridge sample overlaps significantly with these catalogue entries. We further compare with a catalogue of cepheids from \cite{jacyszyn-dobrzeniecka16}. The positions of the classical cepheids in the sample match the neutral hydrogen overdensities, which leads the authors to conclude that they formed in situ after the last interaction between the MCs. The anomalous cepheids of the sample are more spread in their positions. In the LMC outskirts, the ridge line of the NN Bridge sample is offset by $\sim 1^\circ$ from the gas and the cepheids. This could be an indicator that the stars of the NN sample in this area have been stripped from the SMC, not formed in-situ like the cepheids. %Or simply formed at an earlier time, 'using up' the HI in the area? Not as likely, the overall lower HI density near the LMC does not have a corresponding large number of young stars in the area. Maybe young stars found there have travelled there?}
%paragraph on the velocities 

From our previous calculation of the length of the Bridge ($\sim 15$ kpc, see Sect. \ref{subsec:pos}), and the median velocity of the NN Bridge stars travelling away from the Bridge ($\sim114\;\text{km s}^{-1}$, see Sect. \ref{subsec:pm}), we can comment on the feasibility of the stripping scenario. 
Assuming a constant velocity of $\sim114 \; \text{km s}^{-1}$, it would take a star $\sim125$ Myr to cross the $\sim15$ kpc long Bridge in its entirety. Stars that have travelled this distance would have to have been stripped from the SMC at a minimum of 125 Myr ago. Given that the last encounter of the MCs is commonly estimated to have occurred between 100-250 Myr ago, stellar stripping is a plausible scenario to explain the location of young stars in the stellar Bridge. 

%stripping vs. in-situ? Can't draw conclusions!
The alignment of the young stellar population with the neutral hydrogen could point to both tidal stripping of stars and in-situ formation from the tidally stripped gas. 
Studies of the star formation history of the Bridge report enhanced star formation $\sim100$ Myr ago \citep{harris07, noel13, skowron14, ripepi17, Ficara26}. This corresponds to the estimates of the last close interaction between the MCs \citep{besla12, diaz-bekki12, Zivick2019, Choi2022}. Furthermore, the metallicity evolution of the young Bridge population closely matches that of the SMC, suggesting that these stars formed from material originating there \citep{Ficara26}. Studies of star clusters in the region also find that the younger populations follow consistent chemical trends, reinforcing the interpretation that much of the Bridge material was removed from the SMC during the recent interaction between the Clouds \citep{Bica20, Oliveira23, Ficara26}. %Together, these results favour a scenario in which the Bridge formed primarily through tidal stripping of SMC material, with subsequent in-situ star formation occurring within the stripped gas. 
To confirm the origins of the NN Bridge stars through their metallicities and precise ages, spectroscopic follow-up observations are required. The NN Bridge sample therefore presents suitable candidates to perform such future observations.

%-------------------------------------------------------------------

\section{Conclusions}
\label{sec:conclusions} 
In this work, we present a clean catalogue of young candidate Bridge stars in \gaia DR3. % astrometric and photometric data. 
We identified this sample of stars using \gaia astrometric and photometric data and a combination of dimensionality reduction and a neural network classifier. The threshold classification score was chosen to compromise between purity and completeness. We characterise the NN Bridge sample on the basis of its positions, proper motions, line-of-sight velocities, and CMD. Finally, we discuss the implications of the Bridge sample's properties on possible formation scenarios. 
The main findings of this work are the following:

\begin{itemize} 
    \item We provide a new catalogue of 12,864 Bridge stars in \gaia DR3.
    \item We define limits to the NN Bridge sample using a spline fit to the maximum density ridge (see Fig. \ref{fig:pos}) and measure a Bridge length of $\sim 15$ kpc. 
    \item The young NN Bridge sample aligns well spatially with existing samples from the literature (see Fig. \ref{fig:pos}). However, we find an offset of $\sim 1^\circ$ between the NN Bridge and the HI gas and cepheids near the LMC outskirts.
    \item We compute a median tangential velocity of $\sim114\;\text{km s}^{-1}$ for the NN Bridge stars moving from the SMC to the LMC (see Fig. \ref{fig:pm}). At this velocity, it would take a star $\sim125$ Myr to cross the entire Bridge.
    \item While studies of the metallicity and star formation history of the Bridge generally support in-situ star formation, tidal stripping cannot be excluded as a possible formation channel of the Bridge.
\end{itemize}

Upcoming surveys will significantly advance the characterisation of the Bridge. %\gaia DR4 positions and parallaxes are expected to improve by a factor of 1.4, and proper motions even by a factor of 2.7 \citep{gaiadr3summary}. In addition, 4MOST and SDSS-V will provide spectroscopic information. 
The enhanced astrometric precision of \gaia DR4, combined with spectroscopic data from 4MOST \citep{dejong19} and SDSS-V \citep{kollmeier17, kollmeier26}, will enable a more detailed exploration of its kinematics and chemical properties, helping to disentangle its formation mechanisms.

\section*{Data availability}
The final NN Bridge sample, including the classification score of each star, will be available in electronic form at the CDS via anonymous ftp to cdsarc.u-strasbg.fr (130.79.128.5) or via \href{http://cdsweb.u-strasbg.fr/cgi-bin/qcat?J/A+A/}{http://cdsweb.u-strasbg.fr/cgi-bin/qcat?J/A+A/}.

\begin{acknowledgements}
This work has made use of data from the European Space Agency (ESA) mission {\it Gaia} (\url{https://www.cosmos.esa.int/gaia}), processed by the {\it Gaia} Data Processing and Analysis Consortium (DPAC,
\url{https://www.cosmos.esa.int/web/gaia/dpac/consortium}). Funding for the DPAC has been provided by national institutions, in particular the institutions participating in the {\it Gaia} Multilateral Agreement. MS acknowledges funding by the European Union under the Horizon Europe Marie Skłodowska-Curie Actions Doctoral Network grant agreement no. 101072454 @HorizonEU research and innovation programme. OJA acknowledges funding from ``Swedish National Space Agency 2023-00154 David Hobbs The GaiaNIR Mission'' and ``Swedish National Space Agency 2023-00137 David Hobbs The Extended Gaia Mission''. M.R.G. and X.L. acknowledge that this work was (partially) supported by the Spanish MICIN/AEI/10.13039/501100011033 and by "ERDF A way of making Europe" by the European Union through grants PID2021-122842OB-C21 and PID2024-157964OB-C21, the Institute of Cosmos Sciences University of Barcelona (ICCUB, Unidad de Excelencia María de Maeztu) through grant CEX2024-001451-M and the project 2021-SGR-00679 GRC de l’Agència de Gestió d’Ajuts Universitaris i de Recerca (Generalitat de Catalunya).
\end{acknowledgements}

\bibliographystyle{aa} % style aa.bst
\bibliography{bibliography}

\clearpage
\appendix
\section{UMAP sample selection}
\begin{figure}
\includegraphics[width=0.95\columnwidth]{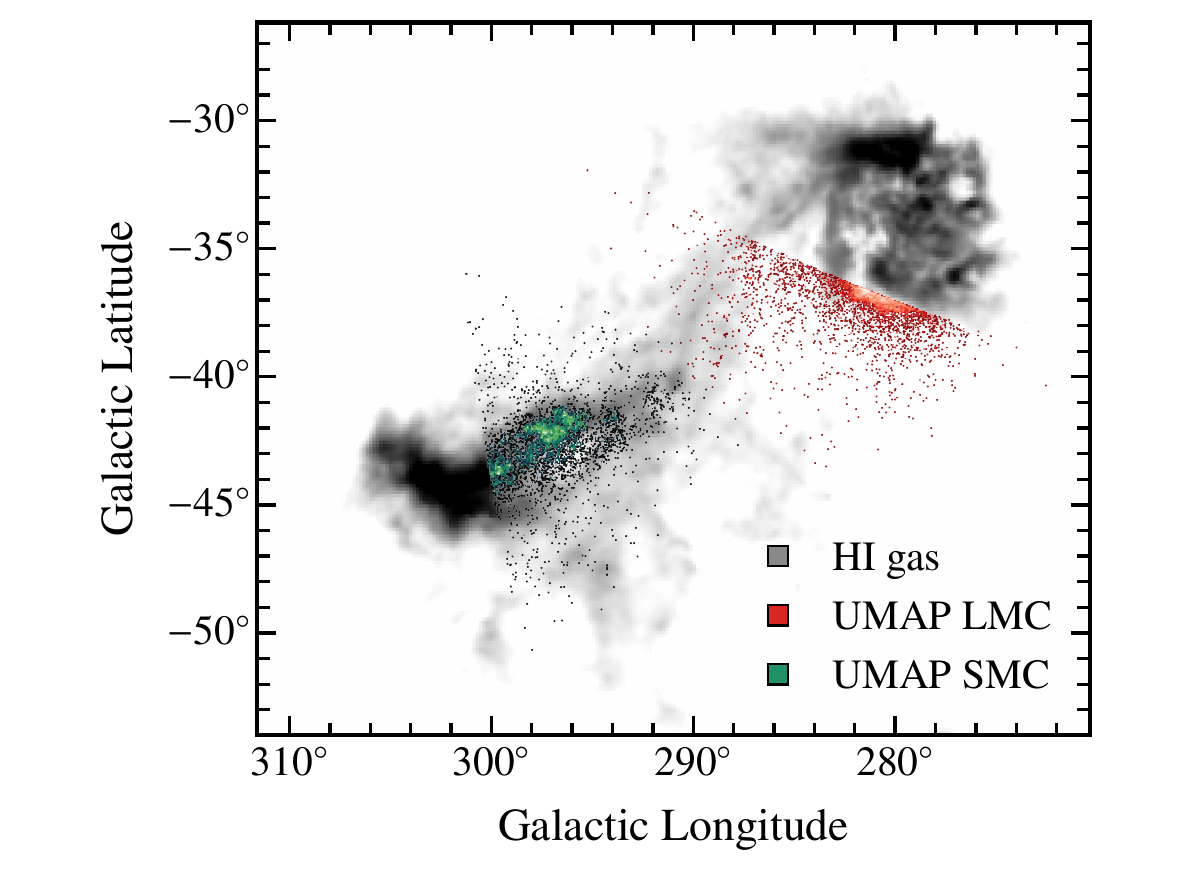}
\caption{Density plot of the \gaia inter-Cloud stars selected from the UMAP feature map (Fig. \ref{fig:umap1}), on top of a neutral hydrogen map from GASS \citep{GASSI, GASSIII}. The red (green) density map shows stars enclosed in the red (green) rectangle, which correspond to young LMC (SMC) stars.
Lighter colours indicate higher density and bins with less than two stars are shown as individual scatter points.}
\label{fig:umap_pos}
\end{figure}

The UMAP feature map (Fig. \ref{fig:umap1}) shows the distribution of the full \gaia inter-Cloud sample in a reduced two-dimensional feature space. By embedding the sample of young LMC and SMC stars in this map, we narrowed down several areas of interest. To evaluate which of these regions correspond to the young stellar Bridge, we investigated their spatial distributions. In Fig. \ref{fig:umap_pos} we show a density plot of the inter-Cloud stars that were selected in the red and green rectangles on the feature map. The green bins show the stars enclosed in the green frame, a population of young stars located in the eastern wing of the SMC. In red, we display the density map of the stars contained in the red rectangle, which represents young stars in the western outskirts of the LMC.

\section{Neural network validation}
\begin{figure}
\centering
\includegraphics[width=0.95\columnwidth]{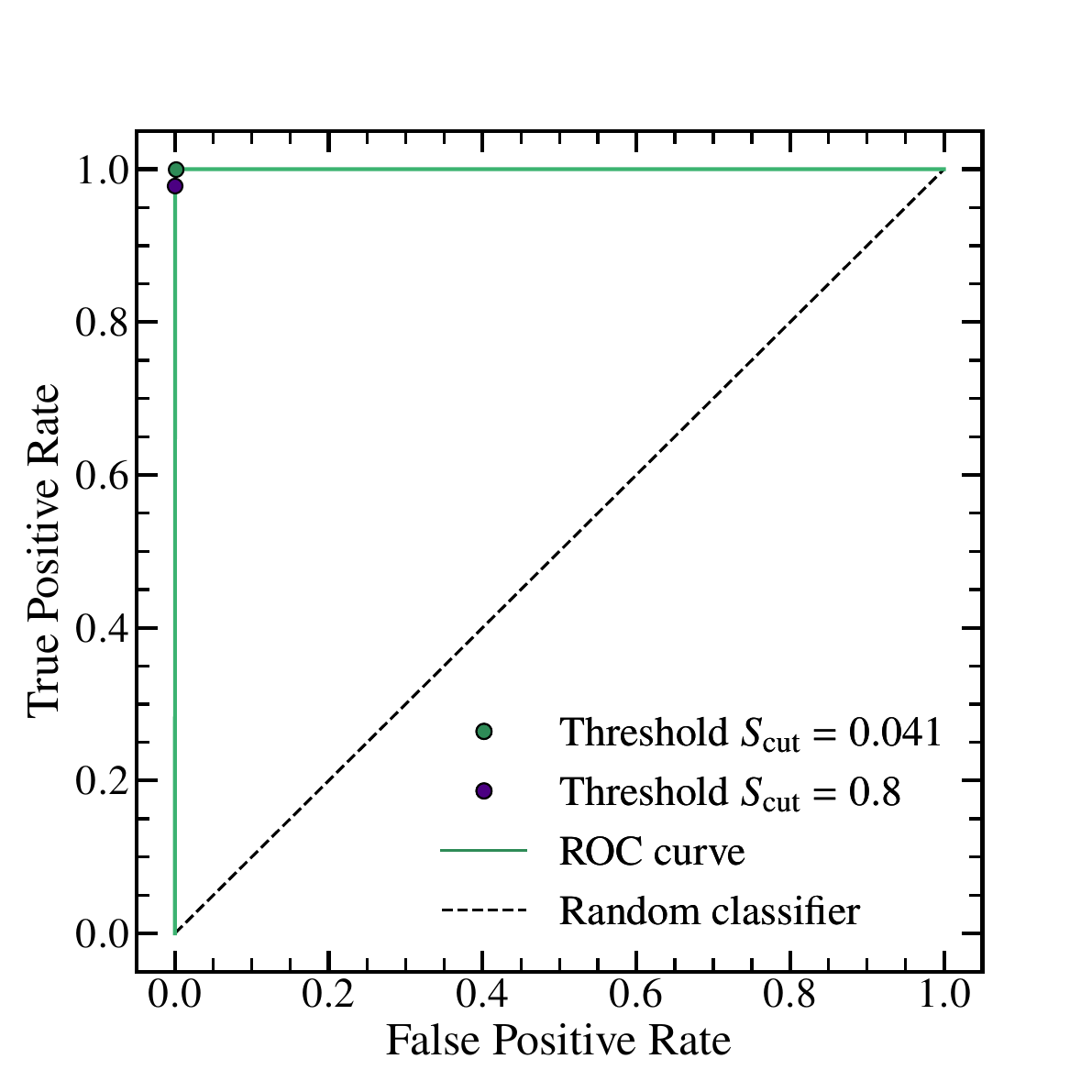}
\includegraphics[width=0.95\columnwidth]{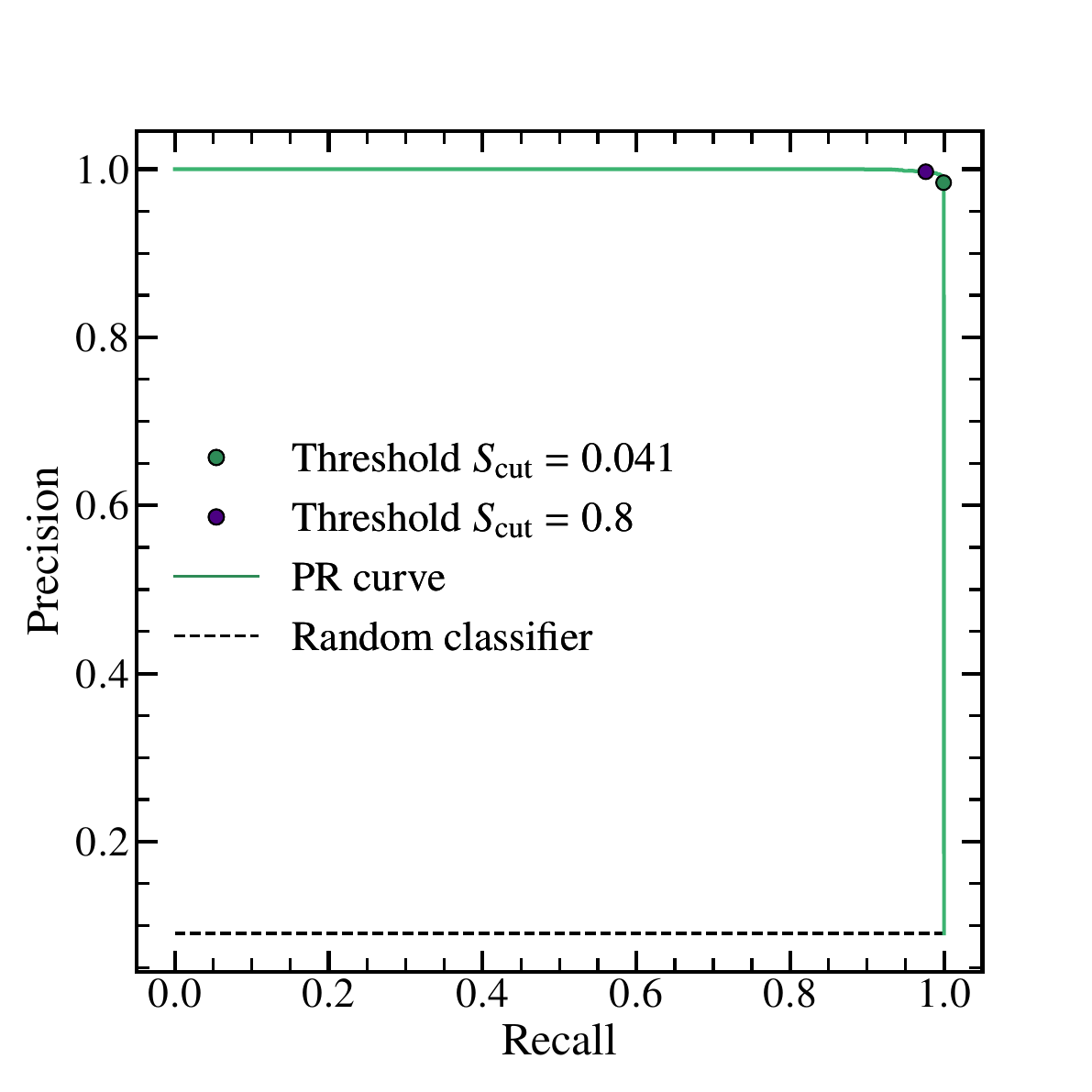}

\caption{Performance curves for the NN classifier (Sect.\ref{subsec:nn}). Top: ROC curve. Bottom: Precision--recall curve. Both curves show the NN performance (solid green lines), compared to a random classifier with no ability to distinguish between classes (dashed black lines). We also plot the threshold $S_\text{cut}=0.041$ (green marker) and the chosen threshold $S_\text{cut}=0.8$ (purple marker).}
\label{fig:roc_pr}
\end{figure}

To evaluate the performance of the NN classifier, we present both the receiver operating characteristic (ROC) curve, and the precision-recall (PR) curve. It is important to note that both evaluation metrics are based on the classifier's performance on the testing dataset. The ROC curve (top panel, Fig.~\ref{fig:roc_pr}) shows the trade-off between the true positive rate and the false positive rate (solid green line) for different classification thresholds, including $S_\text{cut}= 0.041$ and the final sample threshold $S_\text{cut}=0.8$ (green and purple markers, respectively). This illustrates the NN classifier’s ability to discriminate between the two classes, in comparison to a random classifier that cannot separate the classes (dashed black line).

The PR curve (bottom panel, Fig.~\ref{fig:roc_pr}) displays the NN classifier's performance in correctly identifying samples belonging to the positive class while limiting false positive predictions. Again we show the PR curve (solid green line) for different classification thresholds, and highlight $S_\text{cut}= 0.041$ and $S_\text{cut}=0.8$ (green and purple markers, respectively). The PR curve emphasises the performance of the positive class, which is helpful in our case due to the class imbalance of 1:10. The ROC curve and the PR curve both suggest that the NN classifier performs very well.

\begin{figure}
\includegraphics[width=0.95\columnwidth]{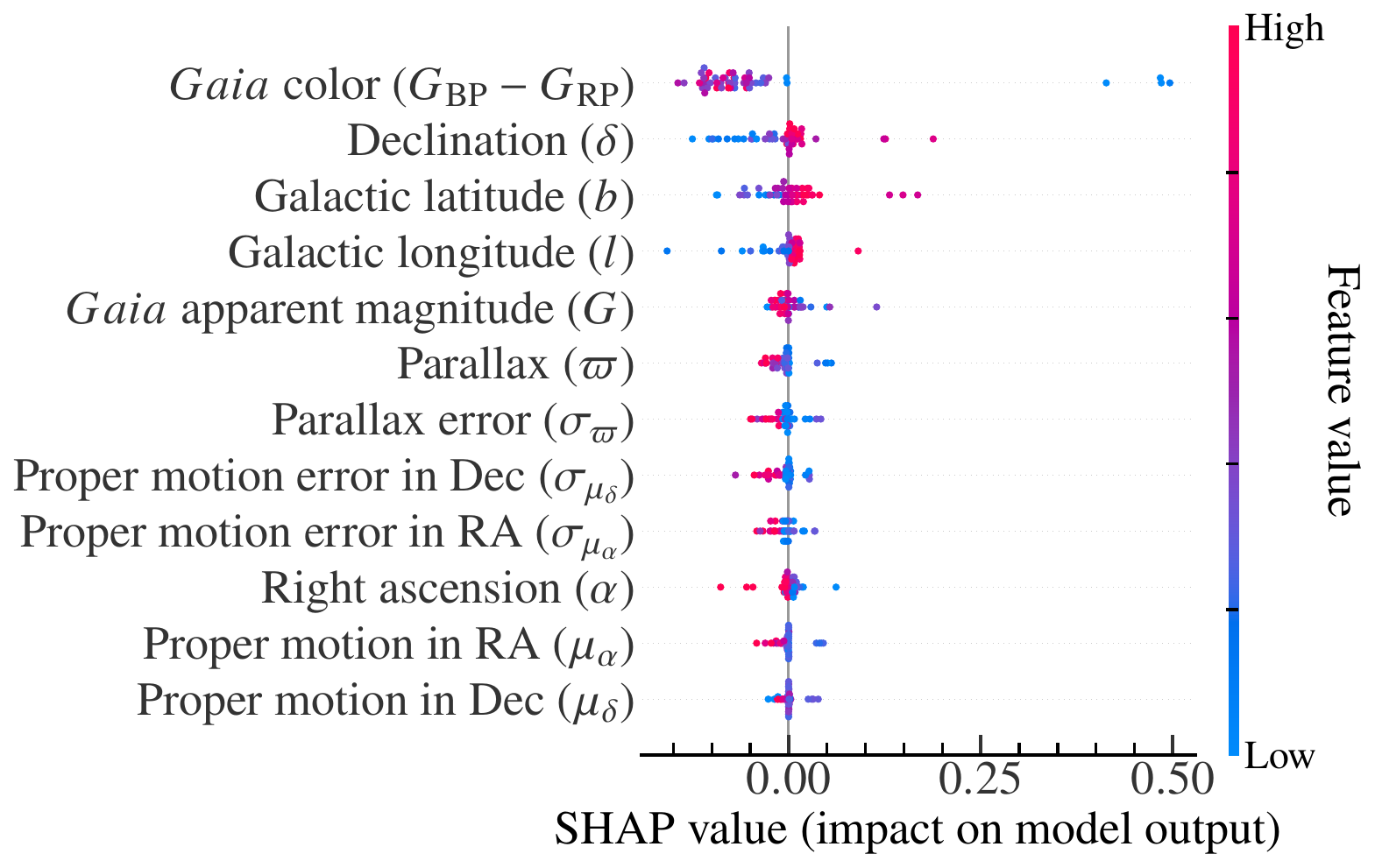}
\caption{SHAP summary plot to visualise how the features from \gaia DR3 astrometric and photometric data contribute to the NN classification of the Bridge-region stars. The parameters used as input to the NN classifier are ordered by impact on the classification. Each data point corresponds to one star in the training set, coloured by its feature value, and positioned according to its proximity to the target class.}
\label{fig:shap}
\end{figure}

Another metric to validate the performance and interpretability of the NN classifier is the SHAP (SHapley Additive exPlanations) plot in Fig.~\ref{fig:shap}. It allows us to quantify how each input feature influences the classifier's predictions and contributes to the final output. The summary plot highlights which features have the strongest impact on the model’s decisions by sorting them in order of their influence on the classification. Every point in the summary plot represents a star of the training sample and has a feature value, where high values are coloured pink and low values blue. The position of each star on the right or the left of the vertical axis indicates how close it is to the $S=1$ or $S=0$ class. The plot thereby visualises how variations in the feature values affect the predicted class probability.

\begin{figure}
\includegraphics[width=0.95\columnwidth]{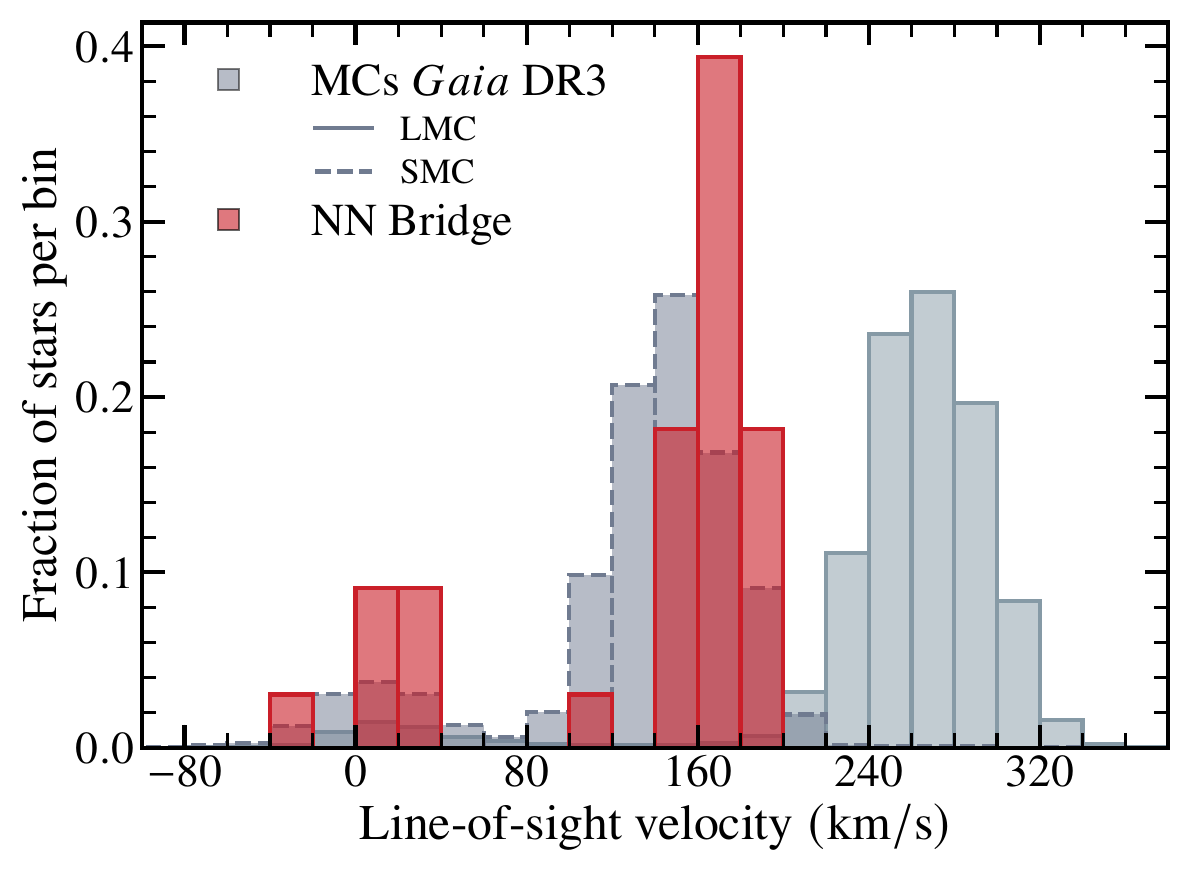}
\caption{Same as Fig. \ref{fig:rv}, but for the NN sample with a classification score threshold of $S_\text{cut}=0.041$}
\label{fig:rv041}
\end{figure}

Lastly, we reiterate the choice of the $S_\text{cut}=0.8$ threshold of the NN sample to achieve high purity while preserving a large number of stars. In Fig.~\ref{fig:rv041}, we show the distribution of line-of-sight velocities of the sample with the $S_\text{cut}=0.041$ threshold suggested by Youden’s J statistic to balance purity and completeness.
Compared to Fig.~\ref{fig:rv}, it includes a higher fraction of stars with line-of-sight velocities in the range between $\sim -40  - 80 \; \text{km s}^{-1}$, which likely corresponds to MW stars. The probable MW contaminants are 7 out of 33 stars for $S_\text{cut}=0.041$, and 2 of 17 stars for $S_\text{cut}=0.8$.

\section{Removing LMC disc stars}
\begin{figure}
\includegraphics[width=0.95\columnwidth]{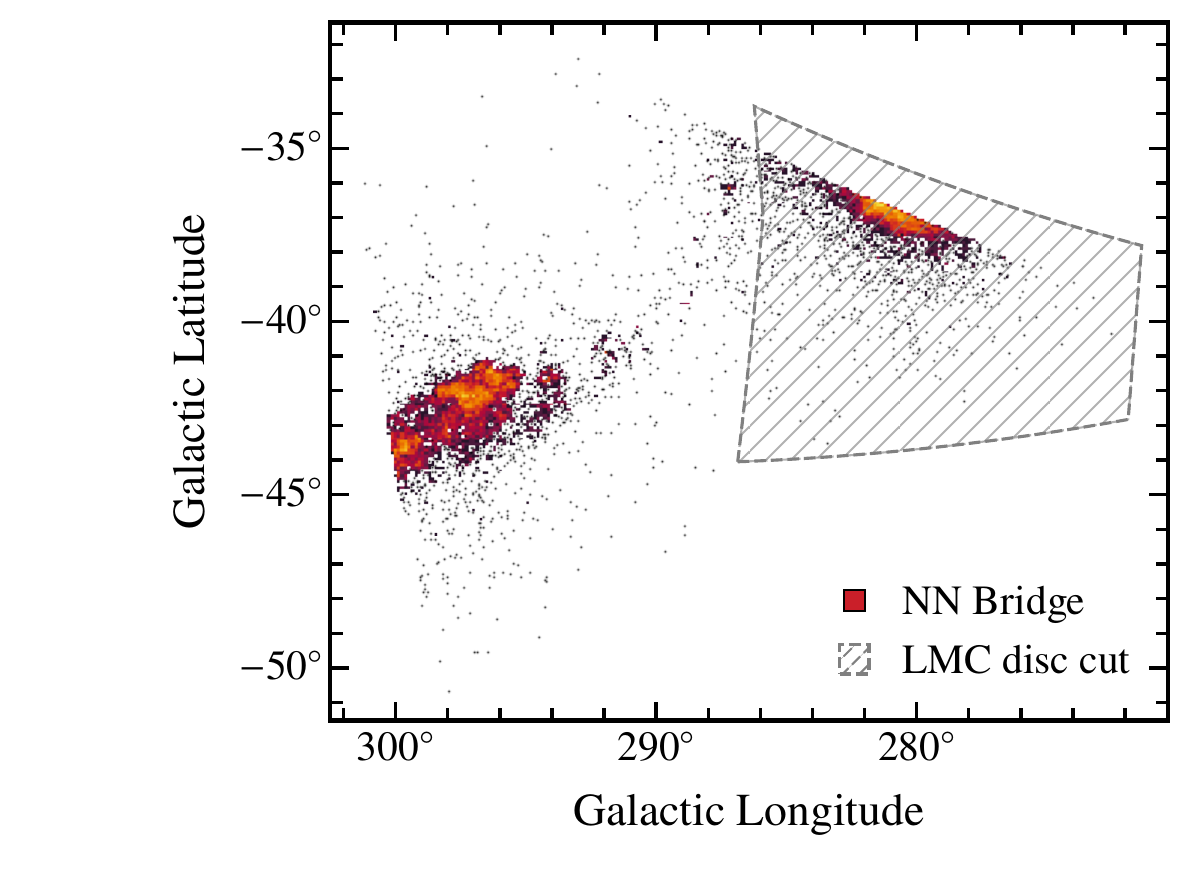}
\caption{Density map of the NN Bridge sample, where bins with only one star are represented as single scatter points. The grey dashed lines and cross-hatched area indicate the positional cut made to remove LMC disc stars from the sample.}
\label{fig:poscut}
\end{figure}

Taking young LMC and SMC stars from the UMAP feature map as the training sample for the NN results in the classification of some stars in the outskirts of the LMC disc as "Bridge" stars. To remove these disc stars, we apply a positional cut, indicated by the polygon in Fig. \ref{fig:poscut} (grey dashed lines and shading). This cut reduces the NN Bridge sample from 19,010 to 12,864 stars. The resulting sample is shown in the middle panel of Fig. \ref{fig:pos} and is characterised in this work.

\end{document}